\documentclass[12pt]{report}

\usepackage{graphicx}
\graphicspath{ {images/} }

\usepackage{color,soul} 
\usepackage{float} 
\usepackage{wrapfig} 

\usepackage{subcaption}

\usepackage[
backend=biber,
style=authoryear,
citestyle=authoryear,
natbib=true
]{biblatex}

\title{The Effects of Climate Change on Predator-Prey Dynamics}

\author[1,*]{Sylvain Gretchko}
\author[2]{Jessa Marley}
\author[3]{Rebecca C. Tyson}
\affil[1,3]{University of British Columbia Okanagan}
\affil[2]{University of Alberta}

\affil[*]{sylvain.gretchko@alumni.ubc.ca}

\date{May 29, 2018}

\begin{abstract}
	In this investigation we study the effects of climate change on predator-prey population dynamics. The interaction of predator and prey is described by the Variable Territory (VT) model with Allee effects in an Ordinary Differential Equation (ODE) framework. Climate influence is  modeled in this ODE framework as an exogenous driver which affects the prey growth rate. Mathematically, this exogenous driver is a function of time, the climate function, that describes how favorable is the climate, on a scale from $-1$ to $+1$. Climate change has a number of effects; in this work, we focus on changes in climate variability, in particular, a decrease in the switching rate between good years and bad.  With regard to the predator-prey populations, we focus on the conditions that lead to extinction. We developed a high performance software tool to allow the systematic exploration of a wide range of climate scenarios at a high level of detail. Our results show that the predator-prey system is sensitive to the rapid drop in growth rate that accompanies a switch from good growth years to bad growth years. The amplitude of this negative variation as well as its duration are key factors, but more importantly, it is the moment when this negative change occurs during the predator-prey cycle that is critical for the survival of the species.
\end{abstract}

\begin{document}

\flushbottom
\maketitle

\thispagestyle{empty}

\section{Introduction}
\label{sec:intro}

Historical climate data shows that climate change is happening. For instance, the evolution of the average temperature in the United States from 1900 to 2015 reveals a clear change in the distribution of temperatures after 1990 \citep{USTemp}. In this example, climate change consists of a clear departure from the longtime averages. There are still yearly fluctuations, but those are now distributed along a different mean value. Another aspect of climate change is the increase in the frequency of extreme climatic events. Such a trend appears for instance in the evolution of extreme one-day precipitation events in the United States from 1910 to 2015 \citep{USHeavyPrecip}.

Little is known about how climate change can affect predator-prey population dynamics, in particular when those populations are cyclic.
\citet{wilmers:2007} explored how climate change and different predation strategies interact to affect aged-structured prey populations. They argued that predators tend to suppress prey population fluctuations when there is a temporal correlation in climatic conditions. They simulated climate by assuming that each year was either good or bad and by controlling the frequency of good years relative to bad years. 

In their paper, \citet{gilg:2009} analyzed the effect of different climate change scenarios on cyclic predator-prey population dynamics in the high arctic. They modeled climate change as variations in the duration of the snow-free period and the frequency of frost-melt events at snow melt. The former affected prey reproduction while the latter introduced random variations in the prey population growth rate. 
They showed that climate change increased the length of the prey population cycle and decreased the maximum population densities. In particular, the latter could negatively affect the predators and could lead to the extinction of some of the four predator species that they studied.

Climate change can also alter the length of summer seasons. \citet{tyson:2016} investigated how this could affect a predator-prey system in which the predator behaved as a generalist predator in the summer, and as a specialist in the winter. They showed that small changes in summer season length could have severe consequences. In particular, increasing summer length could drive the prey to extinction, in contradiction to models where the predator remains a specialist or generalist all year.

In this work, we are particularly interested in the interaction between the periodicity of predator-prey systems and a change in the climatic forcing due to a change in the switching rate between "good" and "bad" years.  In section~\ref{sec:modeling}, we describe the modeling framework and how the notion of climate change is integrated into our model through the concept of a climate function. Simulation experiments and the software platform that made them possible are presented in Section~\ref{sec:exploring}. Following these, we investigate in Section~\ref{sec:CycleChange} the effects of climate change initiated at different points in the predator-prey cycle.  These results lead to further insights into the conditions that lead to extinction. Finally, in Section~\ref{sec:Discussion} we discuss our results and suggest directions for future work.

\section{Modeling Predator-Prey Dynamics and Climate Influence}
\label{sec:modeling}

\subsection{The VT Model}

Predator-prey systems have been studied extensively, and there are many different classical models. We chose as a starting point the Variable Territory (VT) Model (also called the Bazykin model) introduced by \citet{turchin:2001}.

The VT model is defined by a pair of first-order nonlinear differential equations
\begin{subequations}
\begin{align}
	\frac{dN}{dt} = r_0 N \left(1-\frac{N}{K}\right)-\frac{c N P}{d+N} ,      
    \label{eq:VTN} \\
	\frac{dP}{dt} = \chi \frac{c N P}{d + N}-\delta_0 P - \frac{s_0 q}{N}P^2,
\end{align}
\label{eq:VT}
\end{subequations}
where $N$ and $P$ are the densities of prey and predators, respectively. Equations~\eqref{eq:VT} are essentially the Rosenzweig-MacArthur model with an additional density-dependent death term in the predator.  It is this term that is consistent with predator territory sizes increasing as prey density decreases.  A description of the parameters, their units and default values is given in Table~\ref{tbl:VTParameters}. The parameters values were obtained from \citet{strohm:2009} and are based upon data from snowshoe hare and Canada lynx studies in the boreal forest.

\begin{table}
	\centering
	\begin{tabular}{l|l|r|r}
		Parameter & Description & Unit & Default Value\\\hline
		$r_0$ & Intrinsic growth rate for prey& /yr & 1.75\\ 
		$K$ & Carrying capacity for prey& prey & 8\\
		$c$ & Predator saturation kill rate& $\text{prey/(pred}\cdot \text{yr)}$ & 800\\
		$d$ & Predator half-saturation constant &prey & 800\\
		$\chi$ & Prey–predator conversion rate & pred/prey & 0.3\\
		$\delta_0$ & Predator death rate in absence of prey & /yr & 0.0040\\
		$s_0$ & Intrinsic rate of pred. pop. increase & /yr & 0.80\\
		$q$ & Minimum prey biomass per predator& prey/pred & 212\\
		$A$ & Strength of Allee effect for prey& prey & 0.05\\
		$B$ & Strength of Allee effect for predator& pred & 0.000020\\
		$e$ & Climate function scaling factor&  & 0.25
	\end{tabular}
\caption {Description of the parameters used in our modified VT model with their default values. The initial values for prey and predator density are $N_0=3$ prey/ha and $P_0=0.003$ pred/ha respectively.}
\label{tbl:VTParameters}
\end{table}

\subsection{Adding Allee effects}
Since we are interested in the conditions that lead to extinction, we take into account the Allee effects that can be present at low population densities \citep{zhou:2005}.  The model becomes
\begin{subequations}
\begin{align}
\frac{dN}{dt} &= r_0 N \left(1-\frac{N}{K}\right)\left(\frac{N}{A+N}\right)-\frac{cNP}{d+N}, \label{eq:VTalleeN} \\
\frac{dP}{dt} &= \chi \left(\frac{c N P}{d + N}\right)\left(\frac{P}{B+P}\right)-\delta_0 P - \frac{s_0 q}{N}P^2,
\end{align}
\label{eq:VTallee}
\end{subequations}
where the parameters $A$ and $B$ are related to the extent of the $N$ and $P$ ranges, respectively, over which the Allee effect occurs. 
The larger the values of $A$ and $B$, the larger must be the corresponding population in order to escape the Allee effect.

\subsection{Adding the Climate Function}

\begin{figure}
	\centering
	\includegraphics [width=0.5\linewidth] {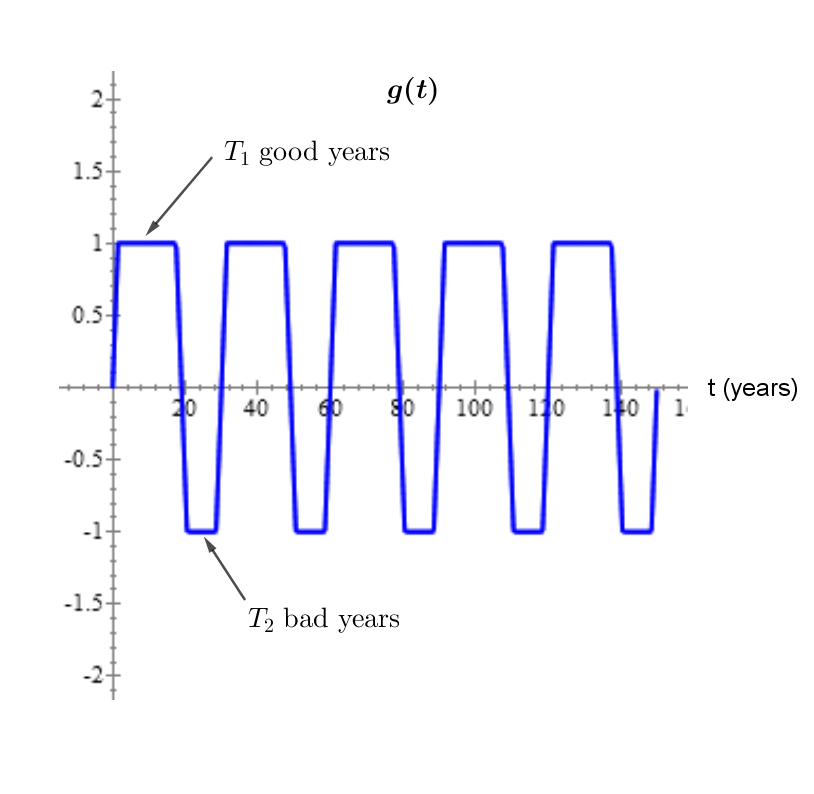}
	\caption {An example of a climate function: an interval of $T_1$ good years is followed by an interval of $T_2$ bad years. Then, the pattern repeats itself, making the function periodic. Transitions between good and bad years are linear with a steepness that can be adjusted.}
	\label{fig:ClimateFunctionExplained}
\end{figure}

Converting the shifting patterns suggested by the historical data into some actual function that can affect the parameters of our model in a meaningful way is not obvious. To reduce this inherent complexity, we will think of the climate influence in terms of good and bad years. From this perspective, climate change can be seen as an evolution in the temporal distribution of good and bad years. We thus define the climate function $g(t)$ as a continuous function of time that has a range of values between $-1$ and $+1$, where $-1$ corresponds to a bad year, and $+1$ corresponds to a good year. Any values between these two extremes are allowed. Figure~\ref{fig:ClimateFunctionExplained} gives an example of a climate function in which intervals of $T_1$ good years alternate with intervals of $T_2$ bad years.
We assumed that the climate function $g$ only affects directly the prey growth rate. This means that predators respond indirectly to climate change through the changes in prey density. This could be achieved by making either the intrinsic growth rate for prey, or the carrying capacity dependent on the climate function. Following \citet{turchin:book}, and temporarily ignoring the Allee effect, we instead chose to affect both parameters in such a way that the slope of the per capita prey growth rate, as a function of prey density, stays constant. That is, the intrinsic growth rate and carrying capacity for prey become
\begin{subequations}
\begin{align}
r_1=r_0\left(1+e\ g(t)\right),\\
K_1=K\left(1+e\ g(t)\right),
\end{align}
\label{eq:rKclimate}
\end{subequations}
where $0<e<1$ is a scaling factor for the climate function.  Using~\eqref{eq:rKclimate} in~\eqref{eq:VTN} we arrive at the following equation for the rate of change of prey density:
\begin{align}
\frac{dN}{dt} &= r_1 N \left(1-\frac{N}{K_1}\right)-\frac{cNP}{d+N}, \\
			  &= r_0 N \left(1+e\ g(t)-\frac{N}{K}\right)-\frac{cNP}{d+N}.
              \label{eq:dNdtclimate}
\end{align}

\begin{figure}
	\centering
	\includegraphics[width=0.5\linewidth]{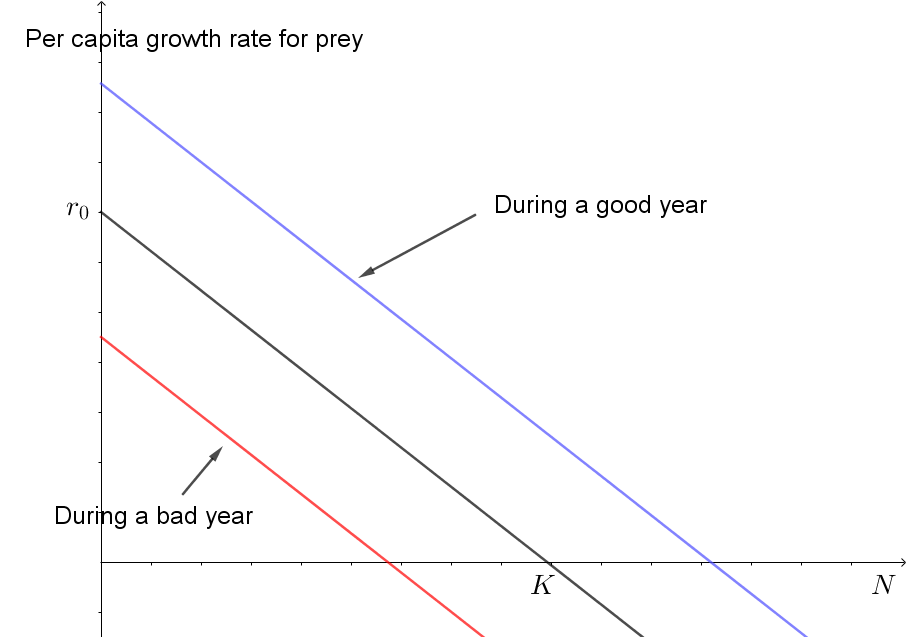} \\
	\caption {Plot of the per capita growth rate as a function of prey density in the absence of predators and Allee effect during a good year (blue line), a bad year (red line) and without climate influence (black line). The black line has intercepts $r_0$ and $K$, corresponding respectively to the intrinsic prey growth rate and carrying capacity for prey.}
	\label{fig:PerCapitaGrowthRate}
\end{figure}

Figure~\ref{fig:PerCapitaGrowthRate} shows how the climate function affects the per capita prey growth rate during good and bad years in the absence of predators and the Allee effect.

Hence, the full VT model with Allee effects and climate function becomes
\begin{subequations}
\begin{align}
\frac{dN}{dt} &= r_0 N \left(1+e\ g(t)-\frac{N}{K}\right)\left(\frac{N}{A+N}\right)-\frac{cNP}{d+N} ,\\
\frac{dP}{dt} &= \chi \left(\frac{c N P}{d + N}\right)\left(\frac{P}{B+P}\right)-\delta_0 P - \frac{s_0 q}{N}P^2.
\end{align}
\label{eq:VTalleeClimate}
\end{subequations}


Without climate function, i.e. when $e=0$, this model exhibits a limit cycle under either good conditions ($g(t)=1$) or bad ($g(t)=-1$), using the default values for the parameters. The main difference between the limit cycles in each case is the period and amplitude of the oscillations.

\section{Exploring the Model's Behavior}
\label{sec:exploring}

\subsection{Development of a Software Platform for Simulations}

While there is some understanding that the switching rate between good and bad years is likely to be altered under climate change, it is not clear what future switching rates we can expect.  We thus wished to perform a detailed exploration of values for $T_1$ and $T_2$. A systematic exploration of this parameter space, however, is a large computational problem. To address this issue a dedicated software was developed in C++ to provide an integrated and interactive environment for the exploration of the model and various climate functions. It is fully multi-threaded and uses the adaptive Runge-Kutta-Fehlberg method \citep{shen:book}. Since it can solve around 10000 systems per second on a quad-core PC, the new software made it possible to explore the parameter space in remarkable detail, revealing the fine-grained structure of the model's response to changes in $T_1$ and $T_2$.

\begin{figure}
	\centering
	\begin{subfigure}[t]{0.45\textwidth}
		\centering
		\includegraphics[width=\textwidth]{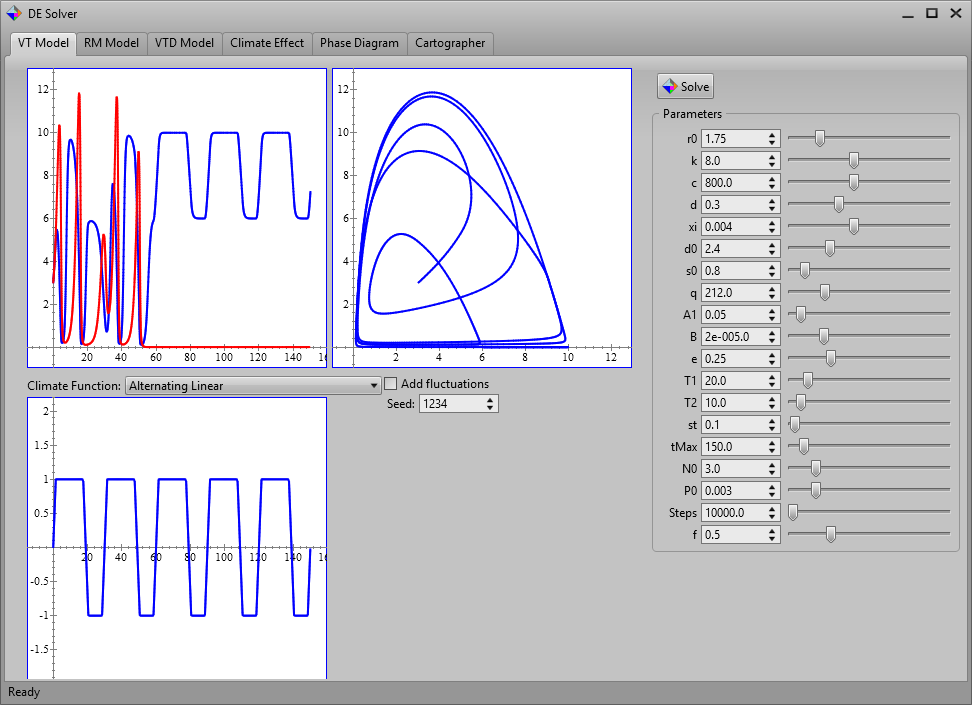}
		\caption {Exploring the predator-prey Model: In this tab, the user can select a climate function and change any parameter of the model to observe in real-time how the solution evolves. It displays both the population densities of prey and predators as a function of time as well as the corresponding phase diagram.}
		\label{fig:SoftwareFistTab}
	\end{subfigure}
	\hfill
	\begin{subfigure}[t]{0.45\textwidth}  
		\centering 
		\includegraphics[width=\textwidth]{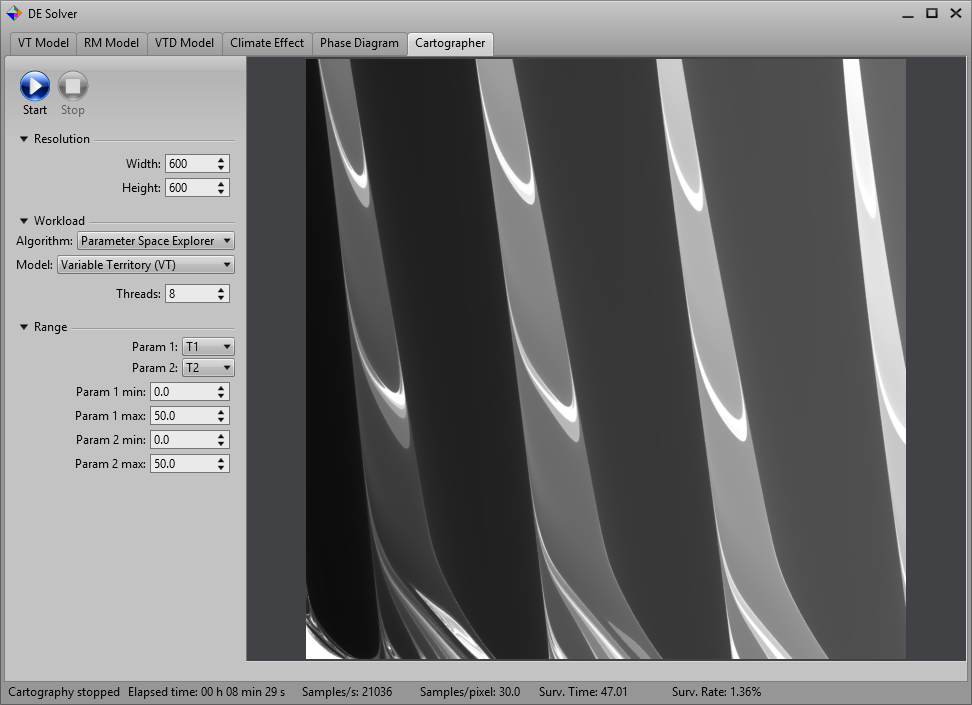}
		\caption {The Cartographer: In this tab, the user can choose two parameters from the model and modify their range. Clicking on the start button begins the progressive rendering of the 2D plot that displays the survival time of the predators.}
		\label{fig:SoftwareCartographer}
	\end{subfigure}
	\caption {Screen captures of the software developed for this work}
	\label {fig:SoftwareScreenCaptures}
\end{figure}

Images of the front-end of the new software are shown in Figure~\ref{fig:SoftwareScreenCaptures}.  The front-end makes it possible to change any parameter of the model and observe in real-time how the solution evolves on the phase diagram and population density plots (Figure~\ref{fig:SoftwareFistTab}). 
%
In addition, an alternate display (the Cartographer) allows the user to fully explore any two-dimensional parameter space of the model (not just the $T_1$ and $T_2$ parameter space) and generate a two-dimensional plot that shows at each coordinate point the survival time of the predator population. A screen capture of the Cartographer is shown in Figure~\ref{fig:SoftwareCartographer}.

The Cartographer uses Quasi-Monte Carlo sampling with the van Der Corput and Sobol low-discrepancy sequences for faster convergence of the generated 2D plot \citep{pharr:2004}. The rendering is progressive and as more and more samples are evaluated the noise in the generated picture gradually disappears. The computation can be stopped at any time, and the resulting image is saved as a Targa file. In this way, it is possible to quickly explore various parameter spaces and get a broad sense of the behaviour of the model in as much detail as desired. 

\subsection{Exploring the Climate Function's Parameter Space}

\begin{figure}
	\centering
	\begin{subfigure}[t]{0.475\textwidth}
		\centering
		\includegraphics[width=\textwidth]{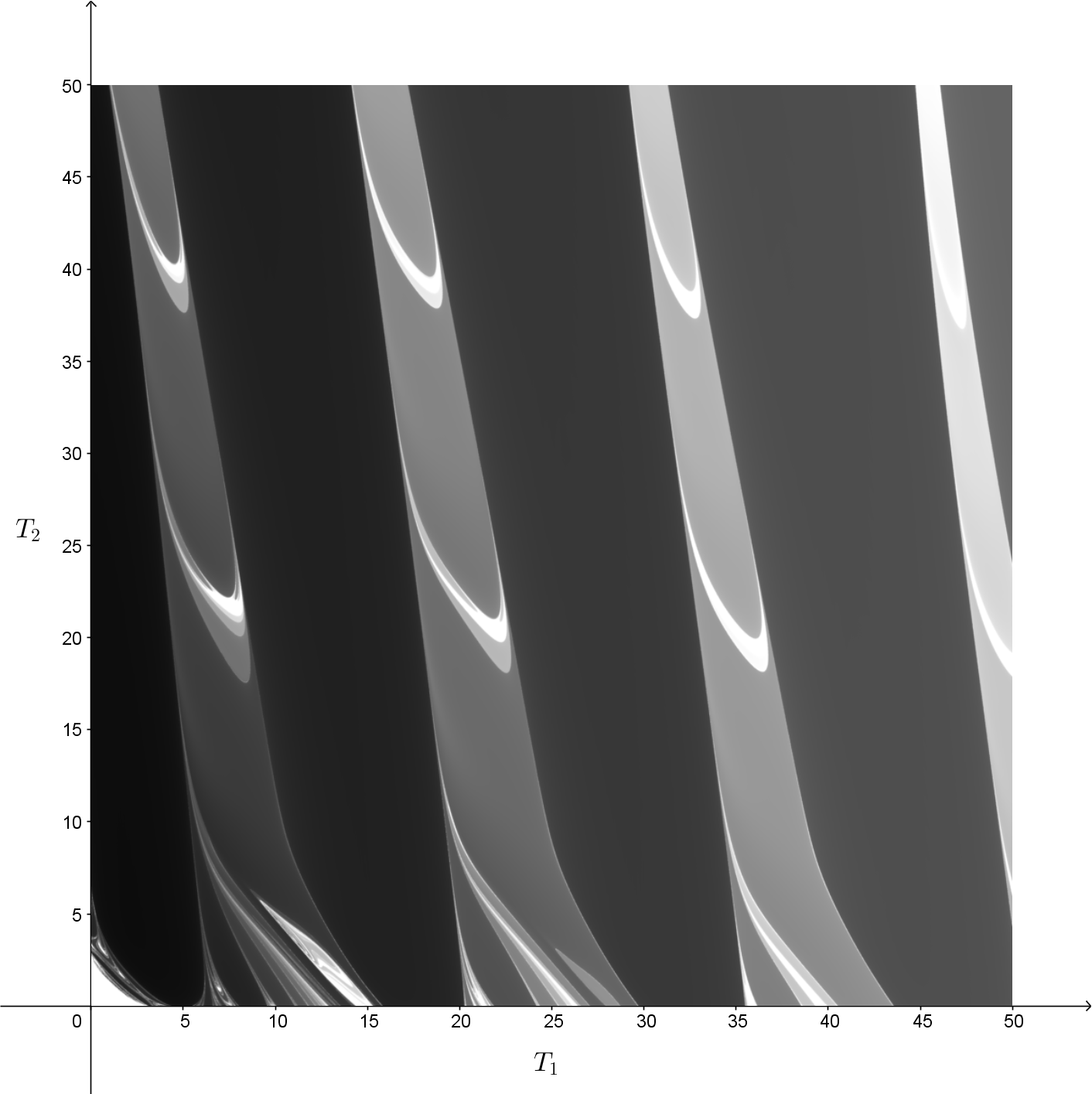}
		\caption {$(T_1,T_2)$ in $[0,50]\times[0,50]$}    
		\label{fig:T1T2ParamSpace_a}
	\end{subfigure}
	\hfill
	\begin{subfigure}[t]{0.475\textwidth}  
		\centering 
		\includegraphics[width=\textwidth]{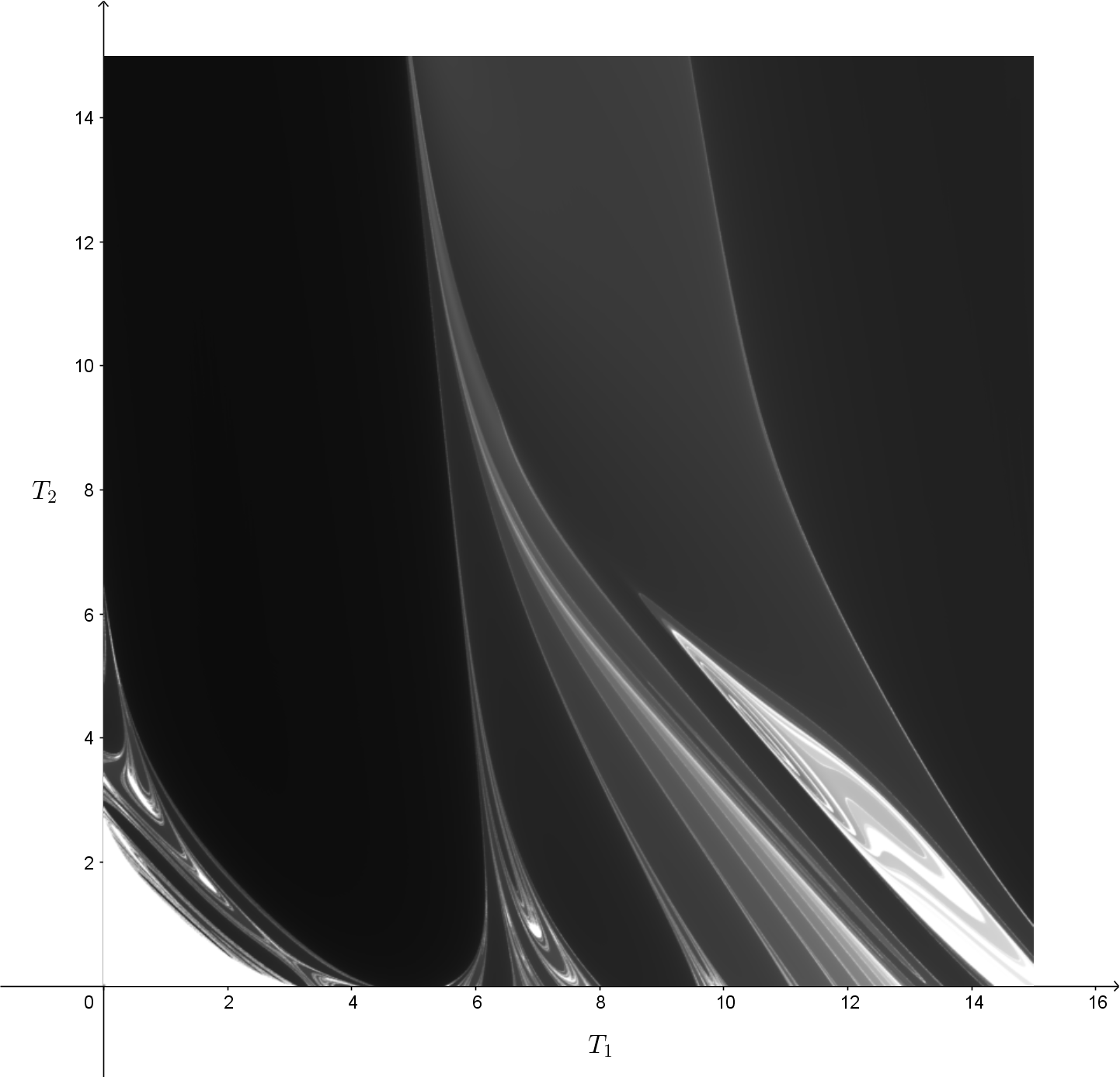}
		\caption {$(T_1,T_2)$ in $[0,15]\times[0,15]$}    
		\label{fig:T1T2ParamSpace_b}
	\end{subfigure}
	\vskip\baselineskip
	\begin{subfigure}[t]{0.475\textwidth}   
		\centering 
		\includegraphics[width=\textwidth]{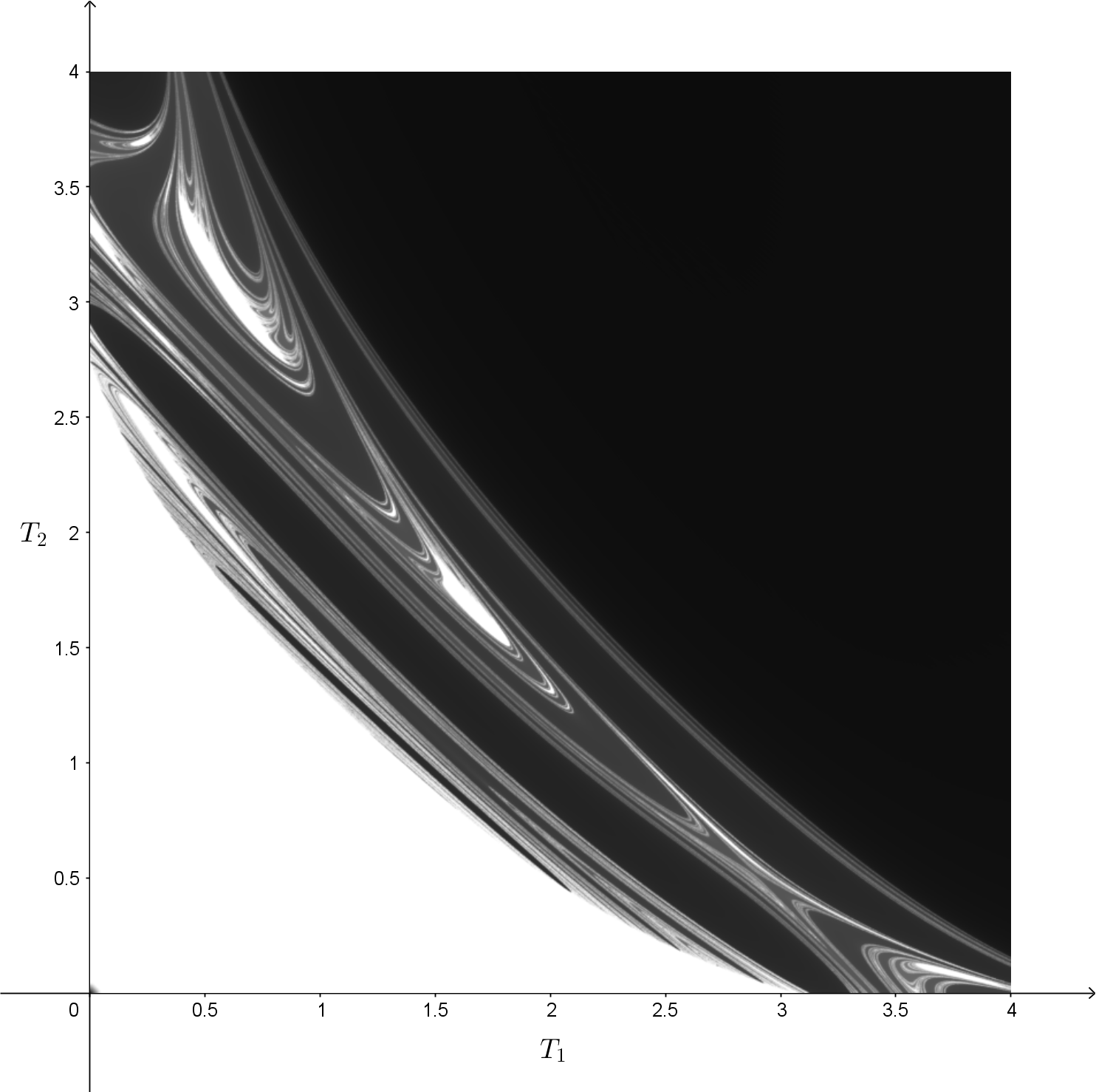}
		\caption {$(T_1,T_2)$ in $[0,4]\times[0,4]$}    
		\label{fig:T1T2ParamSpace_c}
	\end{subfigure}
	\quad
	\begin{subfigure}[t]{0.475\textwidth}   
		\centering 
		\includegraphics[width=\textwidth]{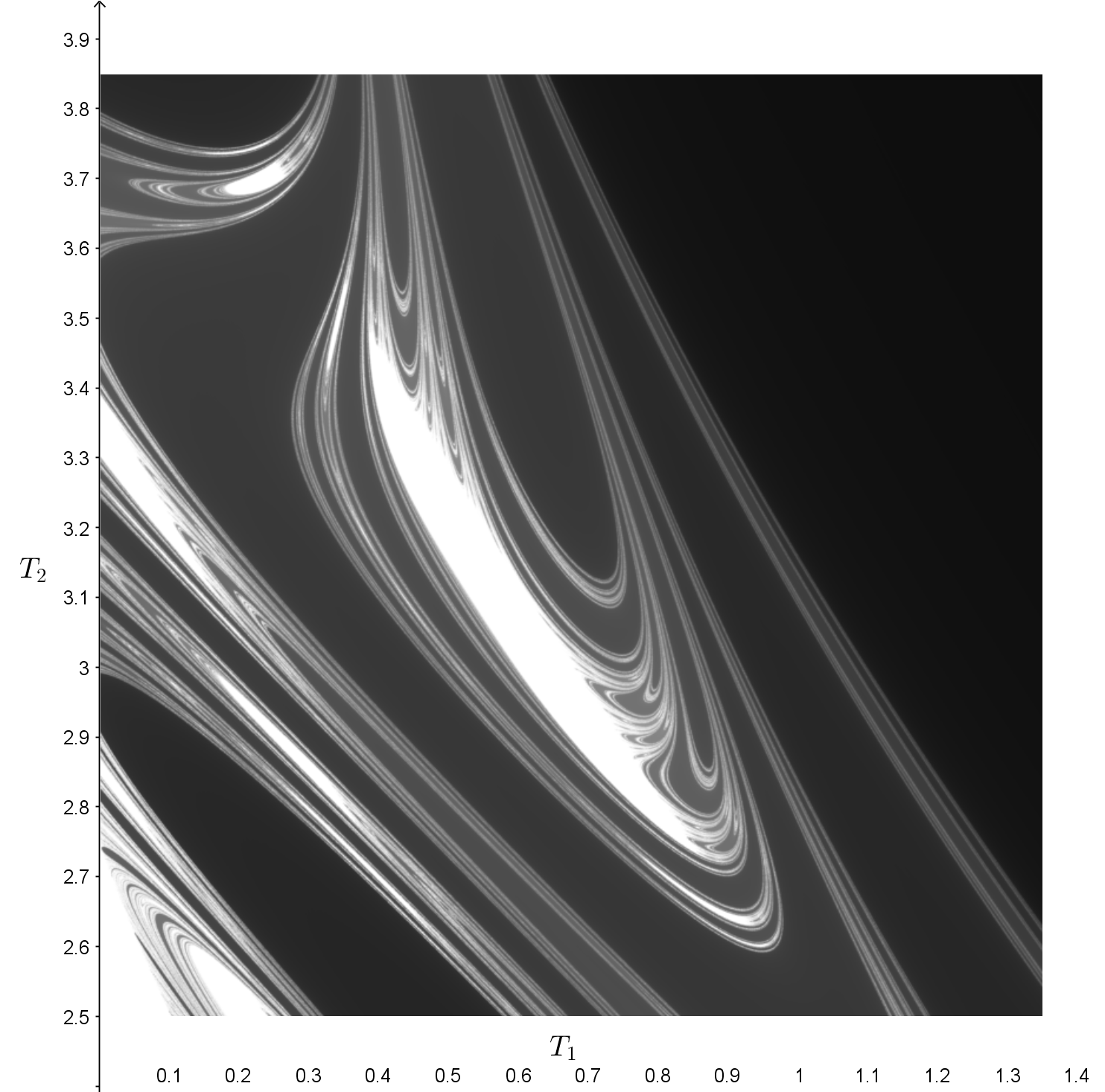}
		\caption {$(T_1,T_2)$ in $[0,1.35]\times[2.50,3.85]$}    
		\label{fig:T1T2ParamSpace_d}
	\end{subfigure}
	\caption {2D plots showing predators' survival time as a function of $T_1$ and $T_2$ using the alternating climate function shown in Figure~\ref{fig:ClimateFunctionExplained}. The plots above were obtained using different ranges, zooming in on details as we move from (a) to (d). Each of these pictures was generated from at least 19 millions of solutions, giving about 30 samples per pixel, and taking between 15 and 40 minutes to render.} 
	\label{fig:T1T2ParamSpace}
\end{figure}

The 2D plot shown in Figure~\ref{fig:T1T2ParamSpace_a} gives the survival time of the predator population for any value of $T_1$ and $T_2$ between 0 and 50 years. We consider only predator survival time, as we found the predator population to be more vulnerable to climate change (as modeled here) than the prey population. In this plot, the color black corresponds to a survival time of zero, while the color white corresponds to the maximal survival time (as defined by the $t_{max}$ parameter which specifies the maximum number of years the simulation will run).  Other survival times appear as shades of grey.  The fine-grained structure apparent in Figure~\ref{fig:T1T2ParamSpace_a} is apparently of a fractal-like nature, as successive plots of smaller and smaller portions of the original plot show similar detail as shown on Figures~\ref{fig:T1T2ParamSpace_b}, \ref{fig:T1T2ParamSpace_c}, and \ref{fig:T1T2ParamSpace_d}.  

We see that the choice of values for $T_1$ and $T_2$ had a dramatic impact on predator survival. The 2D plots reveal a quasi-periodic general structure and intricate sub-structures reminiscent of the two dimensional Poincaré sections of the forced Duffing equation. Over the whole parameter space, where $(T_1,T_2)\in[0,50]\times[0,50]$, the survival rate was only about $1.36\%$, and the average survival time was about 47 years. This means that in only $1.36\%$ of the cases, the populations of prey and predators kept oscillating until the end of the simulation (that is, 150 years after the start). Rate of survival increased for smaller values of $T_1$ and $T_2$. When $T_1$ and $T_2$ were both between 0 and 4 years, the survival rate was $21.32\%$ and average survival time 48.4 years. There was no extinction when $T_2$ and $T_1$ were such that $T_2\leq 2.3-T_1$. This corresponds to the bottom left corner of Figure~\ref{fig:T1T2ParamSpace_c}.

Therefore, fast oscillations of the climate function, that is, a fast switching rate between good and bad years, promote survival. When this switching rate slows down, survival rate drops considerably.

\subsection{Adding Random Fluctuations}

Random fluctuations were added to make the climate function more realistic. To do so, a sequence $r_n$ of random floating point numbers was generated, where $n$ is an integer varying between 0 and the maximum number of years allowed in the simulation ($t_{max}$). Each number $r_n$ is between $-0.5$ and $0.5$. Since the climate function $g(t)$ is defined for any value of the time variable $t$, we compute the perturbation by interpolating linearly the values of the sequence $r_n$. A new parameter $f$ controls the amplitude of the random fluctuations. The resulting perturbed climate function $g'(t)$ is obtained from the unperturbed one $g(t)$ in the following way:
\begin{align}
g'(t)=g(t)+f\ p(t),
\end{align}
where
\begin{align}
p(t) = (1-s)r_n+s\ r_{n+1}
\end{align}
with $n = \left \lfloor{t}\right \rfloor$ and $s=t-\left \lfloor{t}\right \rfloor$.

\begin{figure}
	\begin{tabular}{cccc}
		\includegraphics[width=0.21\linewidth]{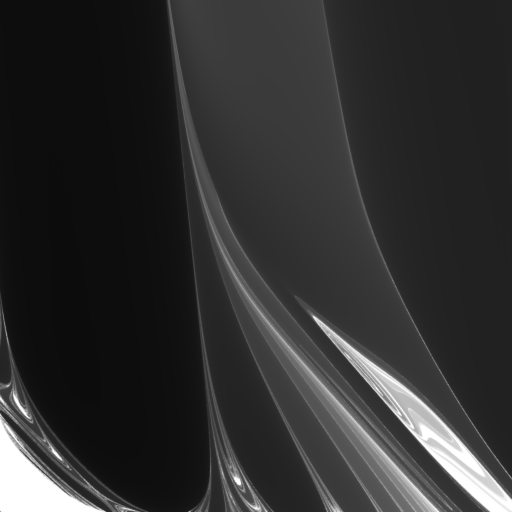}
		&
		\includegraphics[width=0.21\linewidth]{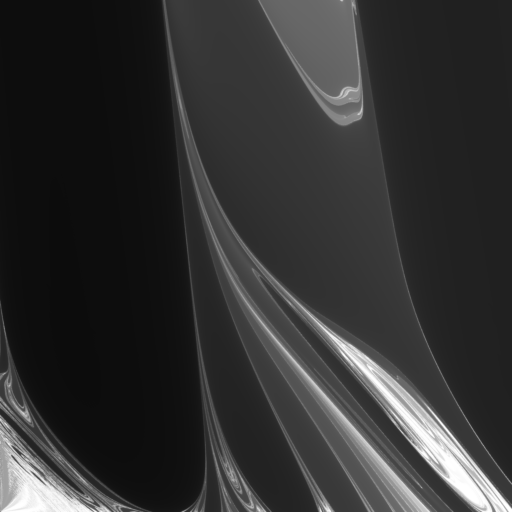}
		& 
		\includegraphics[width=0.21\linewidth]{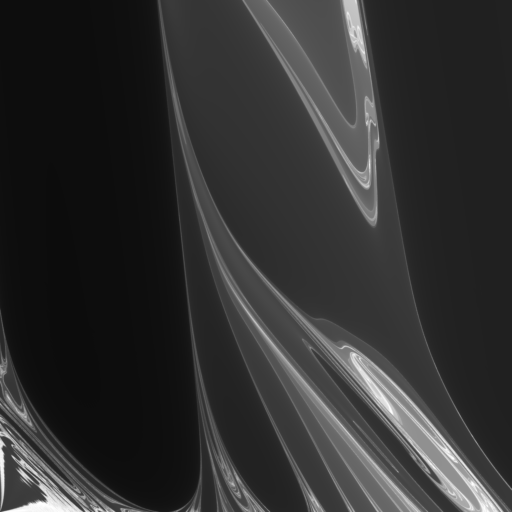}
		&
		\includegraphics[width=0.21\linewidth]{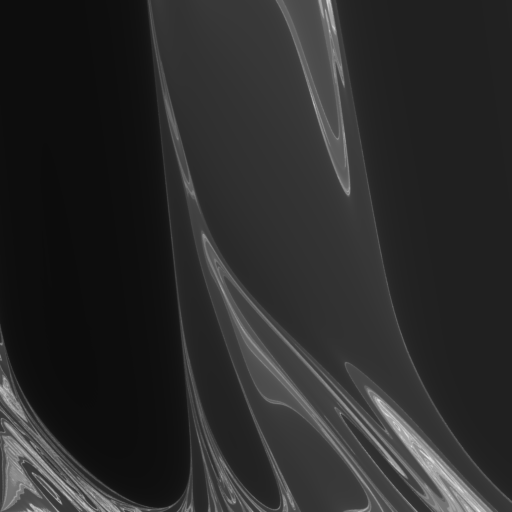} \\ 
		\includegraphics[width=0.2\linewidth]{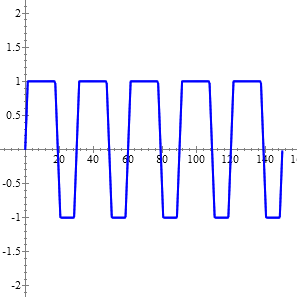}
		&
		\includegraphics[width=0.2\linewidth]{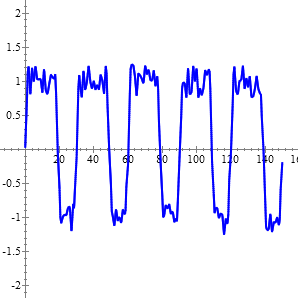}
		& 
		\includegraphics[width=0.2\linewidth]{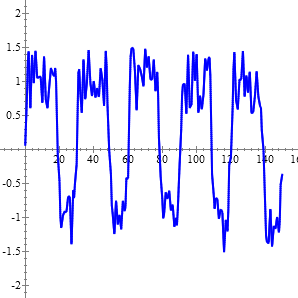}
		& 
		\includegraphics[width=0.2\linewidth]{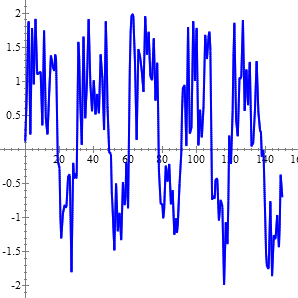} \\ 
		$f=0$ & $f=0.5$ & $f=1.0$ & $f=2.0$
	\end{tabular}
	\caption {Effect of Random fluctuations on an alternating climate function. The top row shows the resulting 2D survival plots for different values of the amplitude $f$ of the random fluctuations. The bottom row shows the fluctuations added to one such climate function.}
	\label{fig:Fluctuations}
\end{figure}

The effect of random fluctuations is shown in Figure~\ref{fig:Fluctuations}. We observe that the general structure of the survival time plot is preserved as the noise level increased, but the survival time drops. In the most extreme case ($f=2.0$), the survival rate dropped to $0.04\%$ for an average survival time of about 25 years.  We conclude that perturbed climate functions have essentially the same effects on the system as the corresponding unperturbed ones, provided that the amplitude of the perturbations remains small relative to the amplitude of the unperturbed climate function. In the remainder of the paper we restrict our attention to unperturbed climate functions.

\subsection{Varying the Amplitude of the Climate Function's Oscillations}

\begin{figure}
	\begin{tabular}{cc}
		\includegraphics[width=0.5\linewidth]{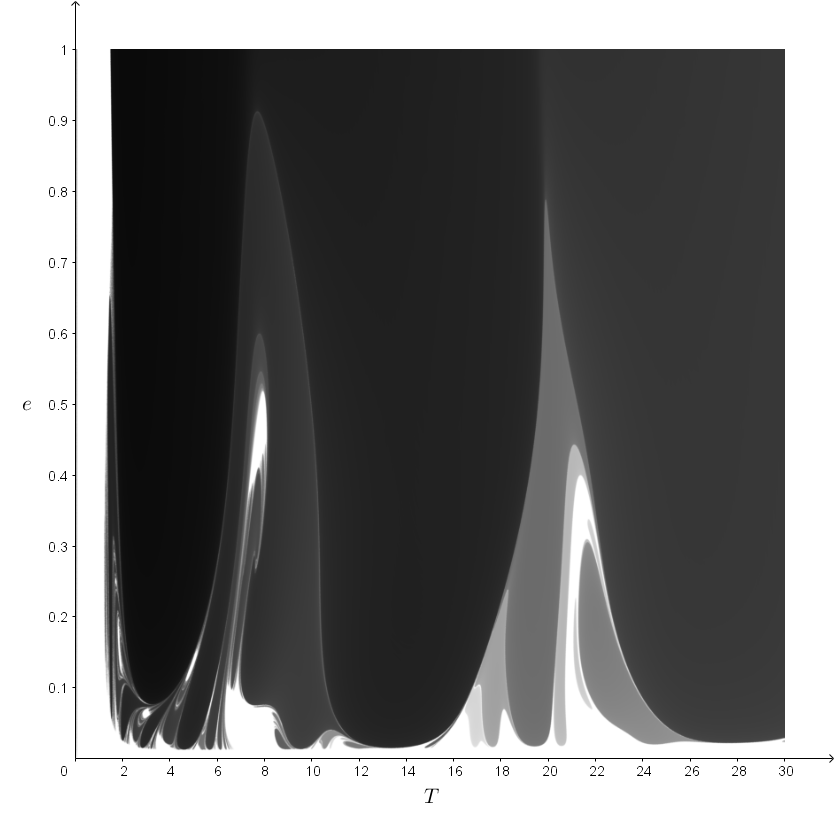}
		&
		\includegraphics[width=0.5\linewidth]{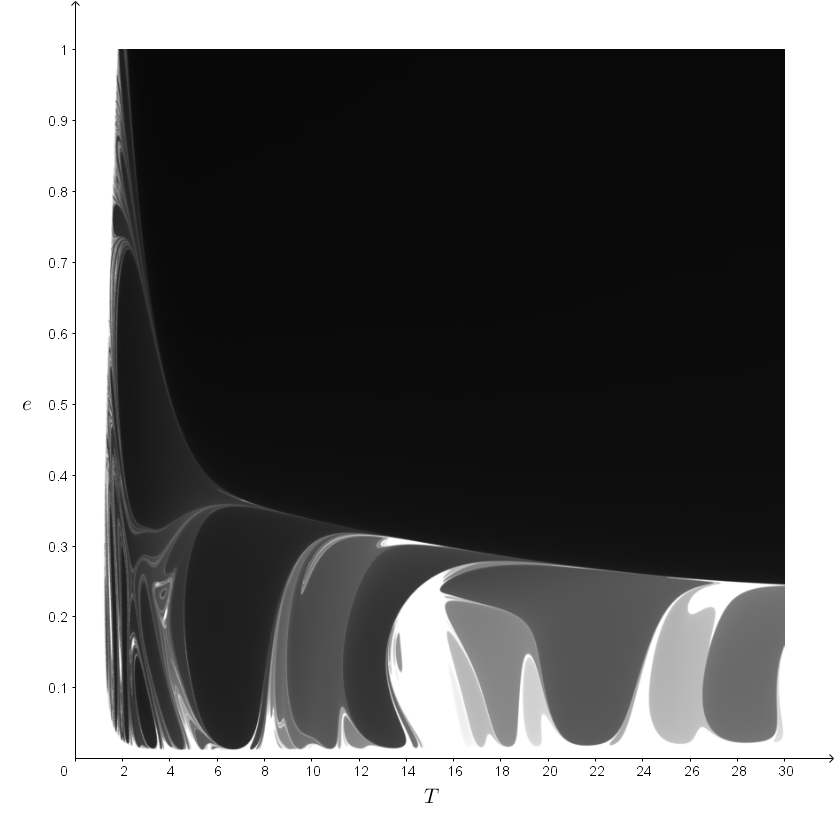}\\
		\includegraphics[width=0.2\linewidth]{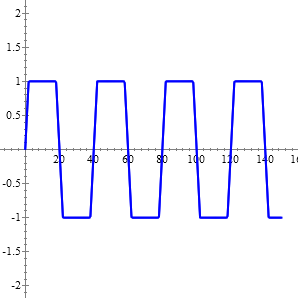}
		&
		\includegraphics[width=0.2\linewidth]{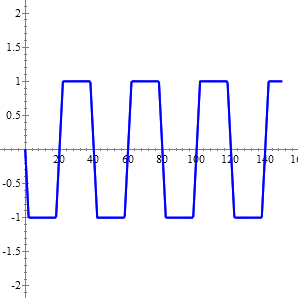}
	\end{tabular}
	\caption {2D plots showing survival time as a function of $T$ (number of good/bad years) and $e$ (amplitude of the climate function) for two simple periodic climate functions (shown below each 2D plot). On the left-hand side, the climate function is defined by $T$ good years followed by $T$ bad years, while on the right-hand side $T$ bad years are followed by $T$ good years.}
	\label{fig:VaryingAmplitudeSingleT}
\end{figure}

Now we consider what happens when we change the amplitude of the climate function, that is, when we vary the scaling factor of the climate function (parameter $e$). 
In order to use the Cartographer, we used a simpler climate function, defined by only one parameter $T$. In this function, shown in the bottom left of Figure~\ref{fig:VaryingAmplitudeSingleT}, $T$ good years are followed by an equal number of bad years (i.e. $T_1=T_2=T$). Using this function and varying both $T$ and $e$, we obtained the 2D plot of the survival time for the predator population shown in the top left of Figure~\ref{fig:VaryingAmplitudeSingleT}. Since starting with good years was arbitrary, a similar 2D map was generated using a climate function for which $T$ bad years are followed by $T$ bad years (see the top right of Figure~\ref{fig:VaryingAmplitudeSingleT}).

Although the patterns were different, both plots shared common characteristics. When $T$ was less than about 1.2 years, there was no extinction, no matter the value of $e$. It was also clear that, except for a few areas, the higher the amplitude, the lower the survival rate. In particular there was no extinction when $e$ was very small, that is, when it was less than $0.01$. Hence, in both cases the amplitude of oscillations tended to affect survival, with survival decreasing as amplitude increased.

\section{The Predator-Prey Cycle and the Effects of Climate Change}
\label{sec:CycleChange}

\subsection{Survival Time as a Function of Phase}

\begin{figure}
	\centering
	\includegraphics[width=0.5\linewidth]{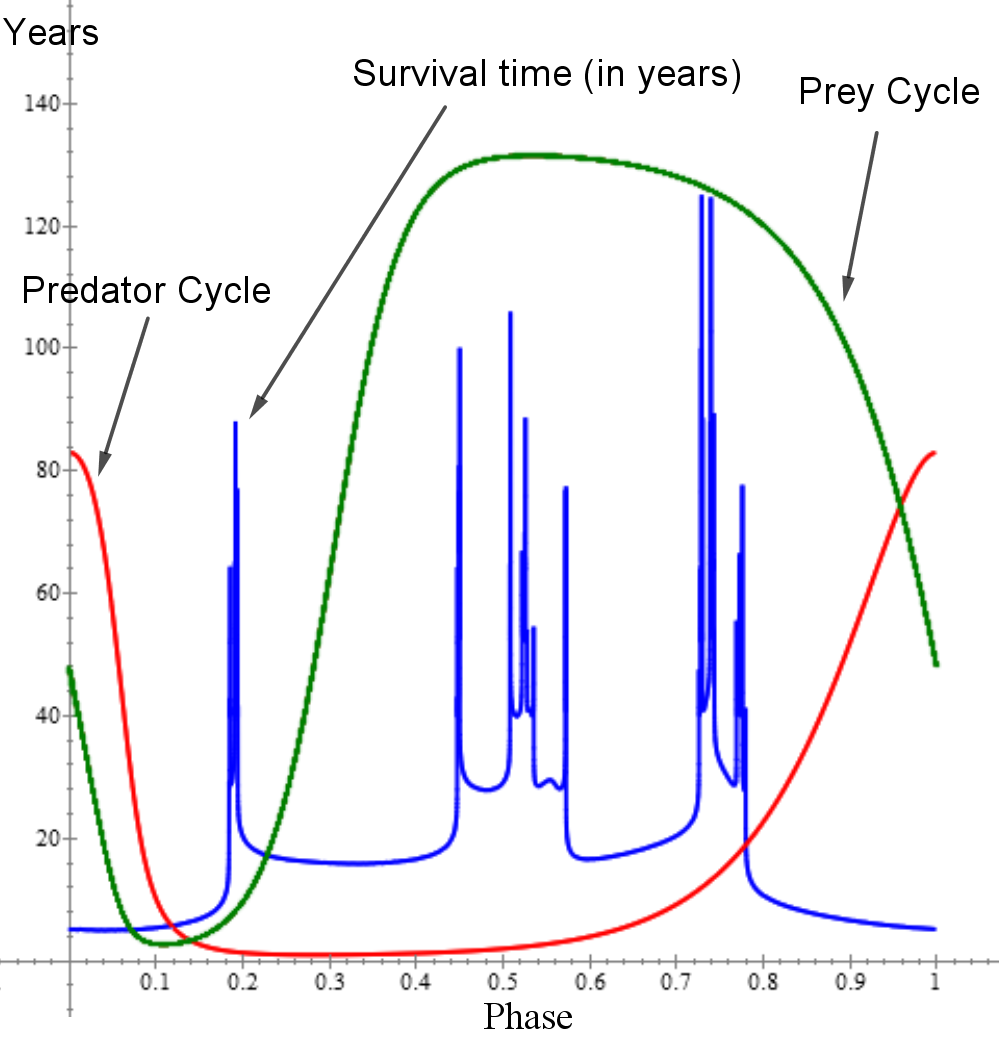} 
	\caption {Plot of predator survival time as a function of the phase (blue curve). The predator and prey cycles are also shown (red and green curves, respectively). The phase is normalized to be between 0 and 1, and represents a specific point in the cycle of both species. A phase of zero corresponds to the maximum of the predator's population. The climate function used here is the same type of simple periodic function shown in the bottom right of Figure~\ref{fig:VaryingAmplitudeSingleT}, with $T_1=T_2=T$, where $T = 6$ years. }
	\label{fig:SurvivalPhasePlot}
\end{figure}

When the climate function is constant, the predator-prey system is self-oscillating. This means that extinction only occurs when the climate function varies. Given earlier work on return maps for perturbation of self-oscillating systems such as heart cells, it is likely that changes in the climate function values will affect the predator-prey system differently depending on the point in the cycle at which the climate changes are initiated. To test this hypothesis, we ran the predator-prey system until a stable limit cycle appeared, and then introduced the climate function at a given point in the cycle. For this simulation, we used the previous climate function with $T=6$ years, starting with bad years. Varying the phase, we obtained a graph of the survival time as a function of the phase (see Figure~\ref{fig:SurvivalPhasePlot}). A phase of zero corresponds to the maximum of the predator population.

The graph shows that survival time is indeed clearly dependent upon the phase, that is, from the moment the climate function starts affecting the predator-prey cycle. Here, predators faced extinction for all values of the phase. The survival time curve has a complex structure; however, survival time appears higher if climate changes are introduced when prey are increasing or when the prey population is high.

\subsection{Varying the Phase and Duration of Good/Bad years}

\begin{figure}
	\centering
	\begin{subfigure}[t]{0.45\textwidth}
		\centering
		\includegraphics[width=\textwidth]{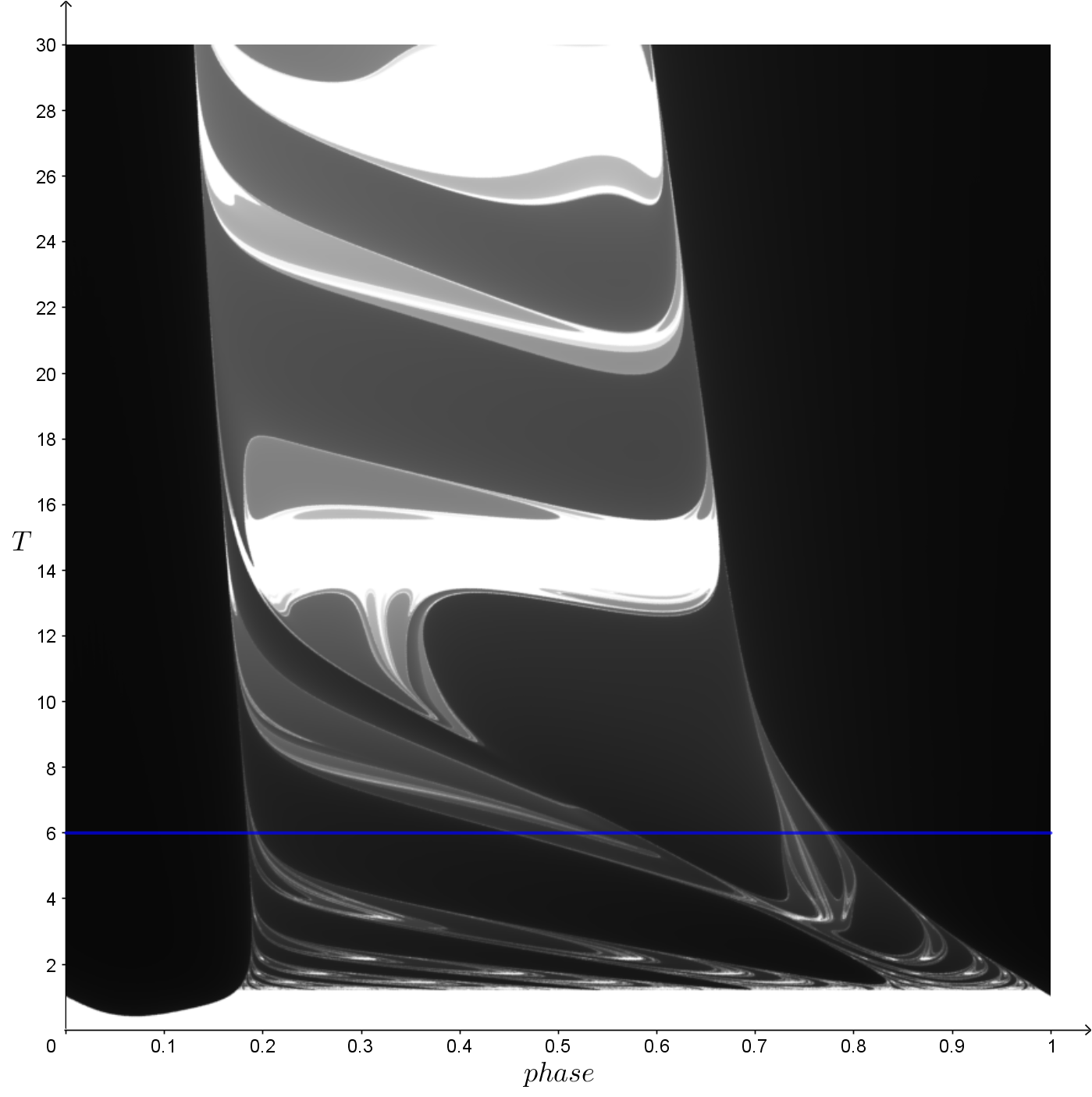}
		\caption {}
		\label{fig:SurvivalPhaseTPlot}
	\end{subfigure}
	\hfill
	\begin{subfigure}[t]{0.45\textwidth}  
		\centering 
		\includegraphics[width=\textwidth]{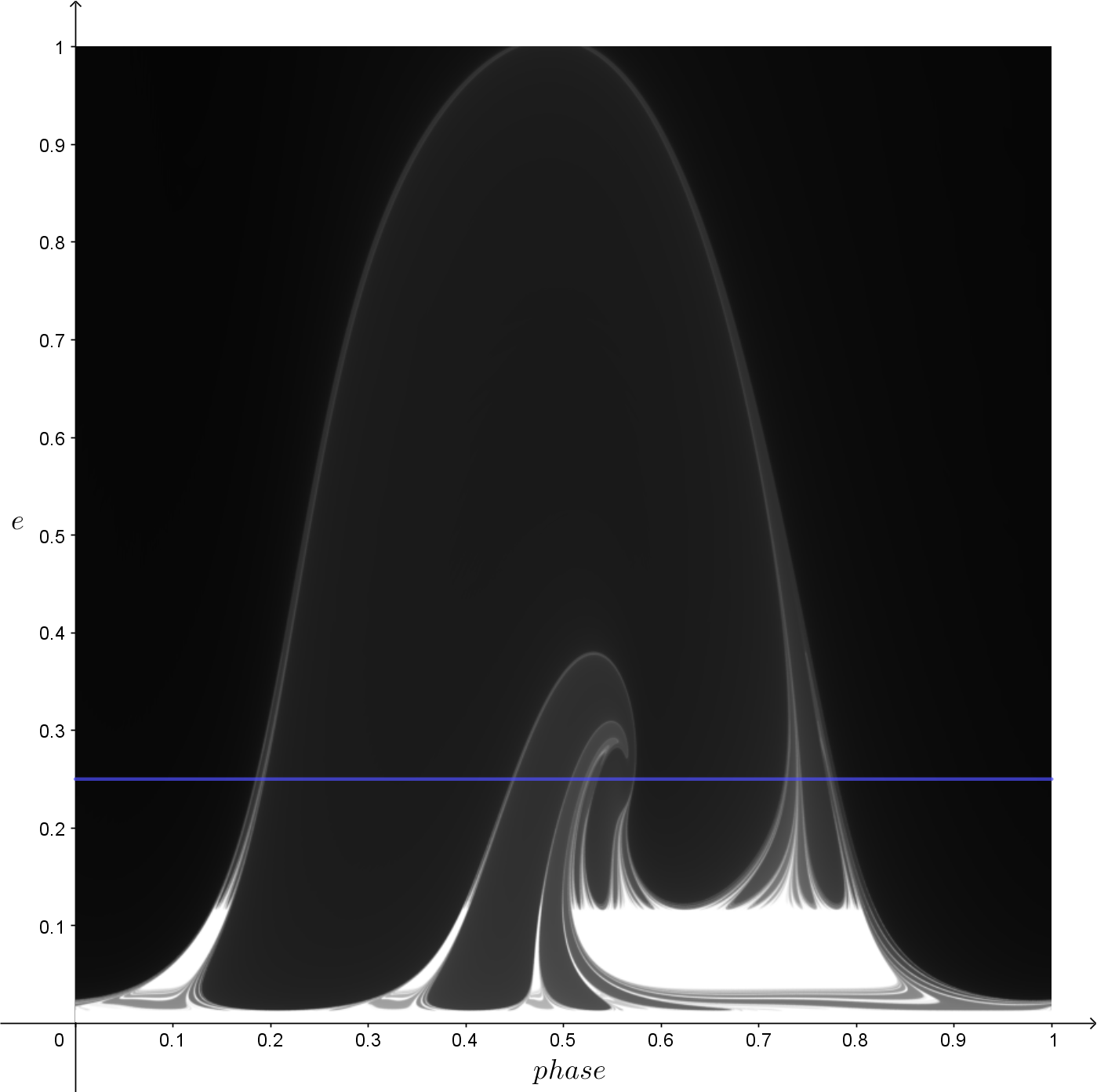}
		\caption {}
		\label{fig:SurvivalPhase_e_Plot}
	\end{subfigure}
	\caption {2D plots showing predators' survival time as a function of the phase and $T$ \textbf{(a)}, and as a function of the phase and $e$ \textbf{(b)}, using the periodic climate function in which $T$ is is the number of bad years, followed by an equal number of good years. $e$ is the amplitude of the climate function. In \textbf{(b)}, we used $T=6$. The blue lines in both plots remind us that the graph in Figure~\ref{fig:SurvivalPhasePlot} corresponds to a slice of these 2D plots. The blue line in \textbf{(a)} has equation $T=6$, while the one in \textbf{(b)} has equation $e=0.25$.}
\end{figure}

Figure~\ref{fig:SurvivalPhasePlot} was created using a single value for $T$, the number of good/bad years in the climate function (starting with bad years). To investigate the effect of changes in $T$, we developed a new tool that could generate a 2D plot showing the survival time as a function of the phase and another parameter chosen by the user. 
The new 2D plot is shown in Figure~\ref{fig:SurvivalPhaseTPlot}, where survival time is shown as a function of phase and $T$. Like the earlier two-parameter figures, it has a complex and fine-grained structure, with regions of very short survival times (black, dark gray) next to regions of long survival times (white, light gray). For small values of $T$ (less than 1.1 years) there is almost no extinction as can be seen in the bottom part of the plot. Otherwise, there is a range of values of the phase for which survival is more likely. This range decreases slightly as $T$ gets larger, but is roughly constant for values of the phase between 0.15 and 0.7, confirming our earlier observation that survival appears higher when prey are increasing or when the prey population is high.

\subsection{Varying the Phase and Amplitude of the Climate Function}

Instead of varying $T$, we are interested here in what happens when we vary $e$, the amplitude of the climate function, along with the phase at which change is initiated. The resulting 2D plot is shown in Figure~\ref{fig:SurvivalPhase_e_Plot}. The amplitude has a strong influence on survival. When $e > 0.11$ extinction becomes prevalent. However, there exists a large area where survival occurs at the bottom right of the plot.  

\subsection{Simplifying the Climate function}

\begin{figure}
	\centering
	\includegraphics[width=0.2\linewidth]{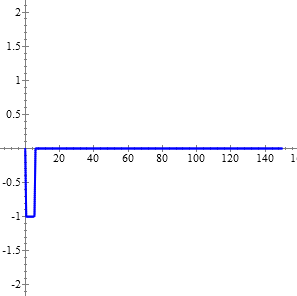}
	\caption {Plot of a climate function consisting of a single negative bump. The duration of the negative bump is $T=6$ years.}
	\label{fig:Climate Neg Bump}	
\end{figure}

Our work suggests that the periodicity of the climate function is responsible for the complex behavior of the system. This leads us to ask how the system would respond to a simple climate function which only exhibits a single negative bump with duration parametrized by $T$. Figure~\ref{fig:Climate Neg Bump} shows the graph of this function for $T=6$. 
Our hypothesis is that a negative change in the climate function at a critical moment in the prey cycle is responsible for extinction. 

\begin{figure}
	\centering
	\includegraphics[width=0.4\linewidth]{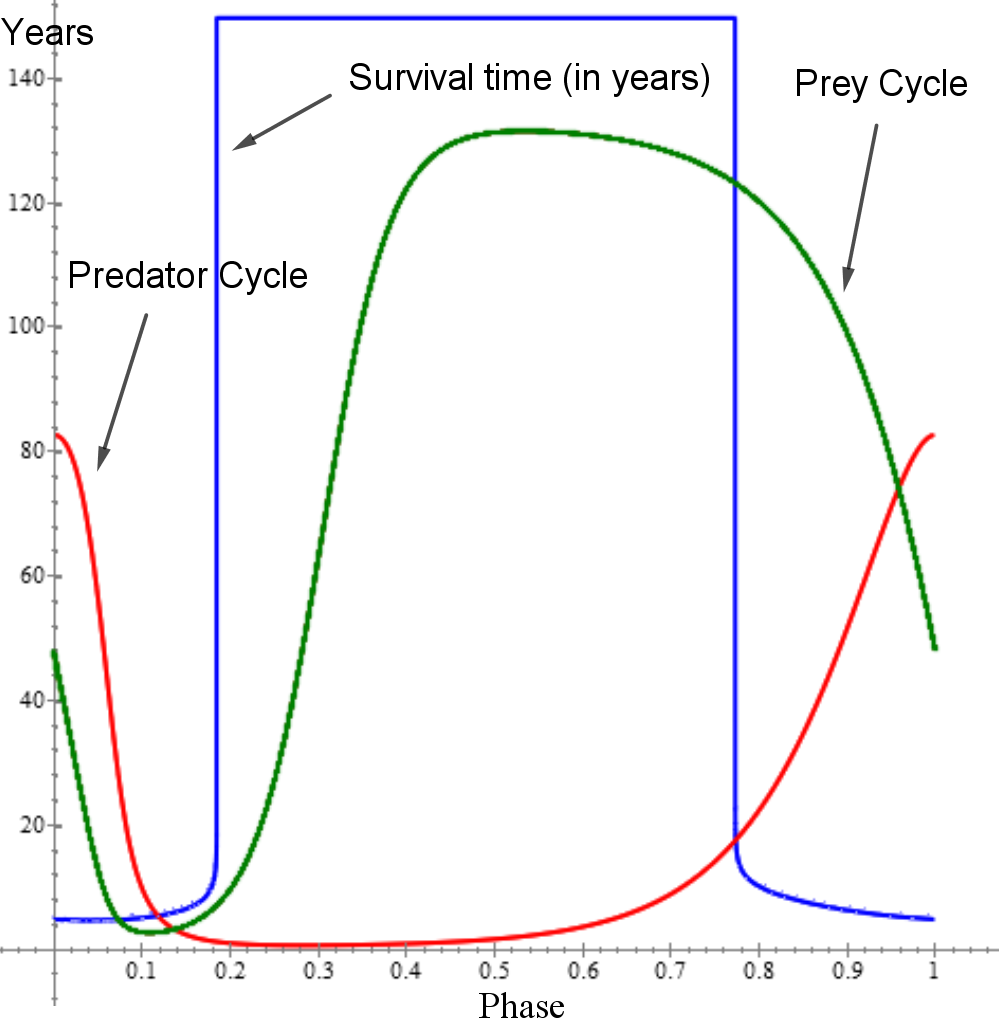}
	\caption {Plot of predator survival time as a function of the phase (blue curve) using a climate function with a single negative bump. The predator and prey cycles are also shown (red and green curves, respectively).  }
	\label{fig:NegBumpSurvivalPhasePlot}	
\end{figure}

Using this simple climate function, we repeated the simulations carried out above. The one-dimensional result is shown in Figure~\ref{fig:NegBumpSurvivalPhasePlot}.  If we compare this plot with Figure~\ref{fig:SurvivalPhasePlot}, we observe that the survival curve for the one-bump climate function is approximately the envelope of the periodic climate function. 
The single bump climate function thus shapes the broad pattern of survival times, while the periodicity adds further detail.  The pattern we see in Figure~\ref{fig:NegBumpSurvivalPhasePlot} confirms our earlier observation that survival occurs if the onset of bad years occurs when the prey density is increasing or close to its maximum.

\begin{figure}
\begin{tabular}{ccc}
	\includegraphics[width=0.19\linewidth]{"ClimateNegBump"}
	&
	\includegraphics[width=0.35\linewidth]{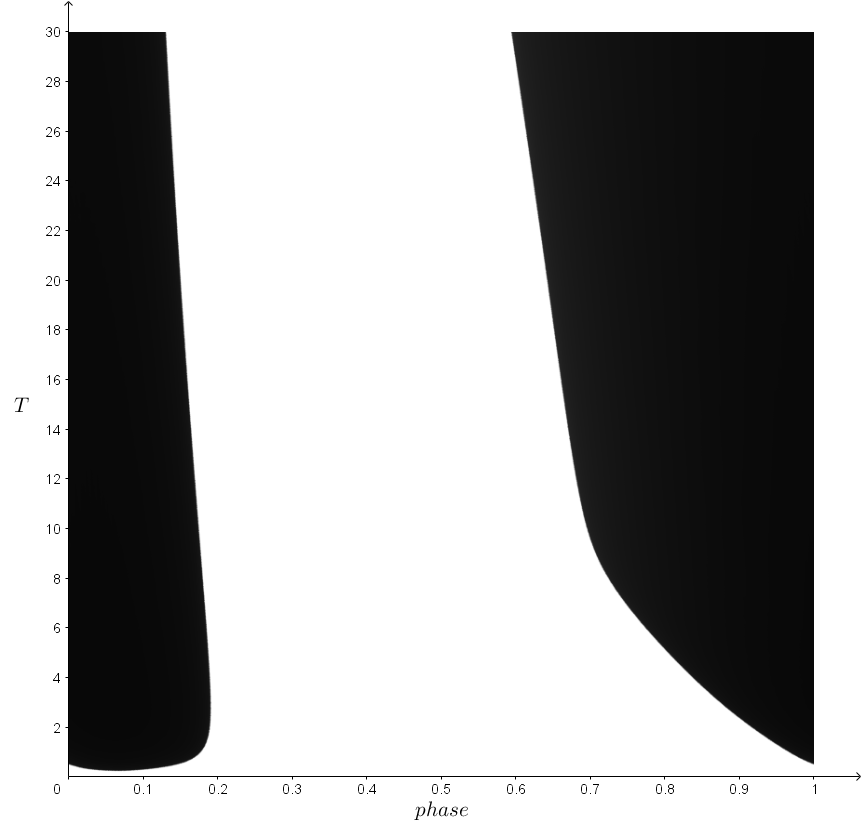}
	&
	\includegraphics[width=0.35\linewidth]{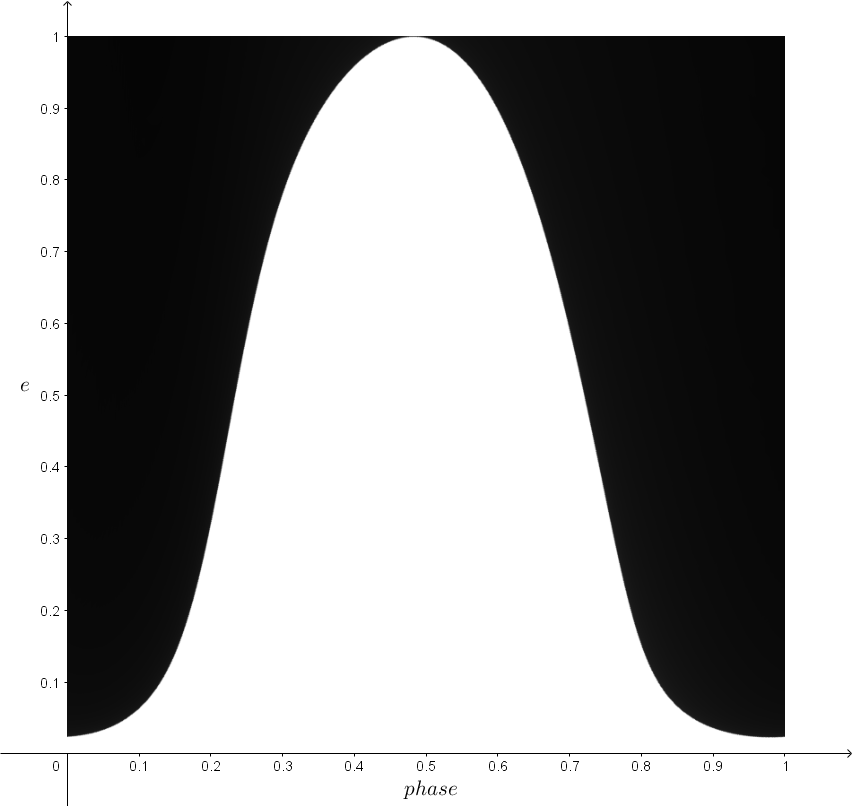}
	\\ 
	\includegraphics[width=0.19\linewidth]{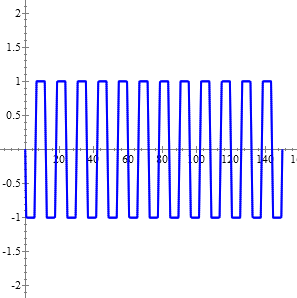}
	&
	\includegraphics[width=0.35\linewidth]{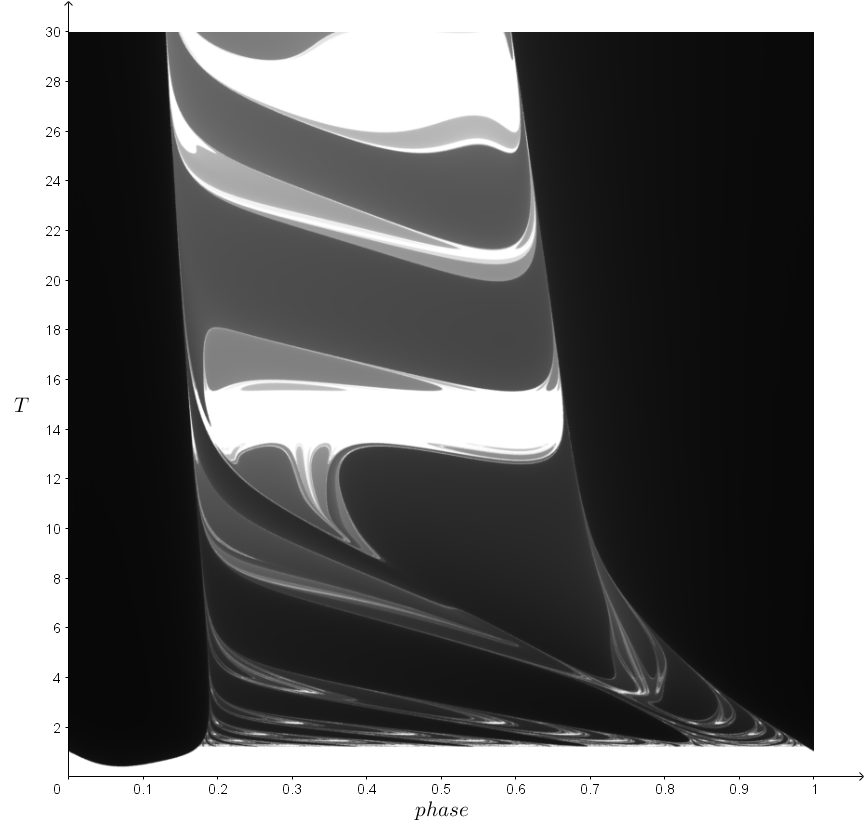}
	&
	\includegraphics[width=0.35\linewidth]{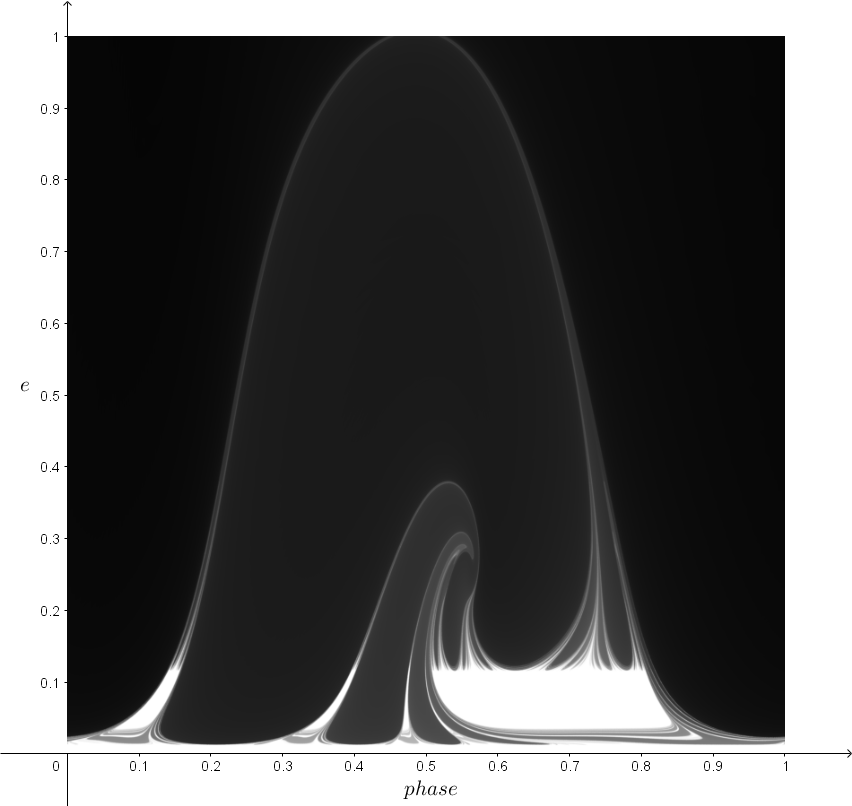}
	\\
	Climate function & $(phase,T)$ & $(phase,e)$
\end{tabular}
\caption {Comparing the effects of two climate functions on survival time. The top row shows the 2D plots of predator survival time as a function of the phase and duration $T$ (middle), and as a function of the phase and amplitude $e$ (right), for the climate function that has a single negative bump lasting 6 years (left). Corresponding 2D plots for a periodic climate function in which $T$ bad years are followed by $T$ good years are given in the bottom row.}
\label{fig:2DSurvivalPhasePlots}
\end{figure}

The effect of the simple climate function in two dimensional parameter space (Figure~\ref{fig:2DSurvivalPhasePlots}, right hand plots, top row) reveals the same type of behaviour: The single bump climate function generates the envelope of the region where survival windows are found with the periodic climate function (Figure~\ref{fig:2DSurvivalPhasePlots}, right hand plots, bottom row).  


The climate function with a single negative bump highlights how the initial bump, or string of bad years, defined by duration $T$ and amplitude $e$, interacts with the periodic switching between good and bad years. By repeating negative bumps at different points of the predator-prey cycle, the periodic climate function increases the likelihood of extinction, unless $T$ and $e$ are in a very specific range.

\subsection{Phase Diagrams}

We can obtain further insights into the system's behaviour by considering phase plane plots for~\eqref{eq:VTalleeClimate}. When there is no climate function, $e=0$ and the system of ODEs is autonomous; the phase plane vector field is unvarying in time. For $e>0$ however, the vector field changes with time.  Another tool was developed to explore this situation. 

\begin{figure}
	\centering
	\begin{subfigure}[t]{0.60\textwidth}
		\centering
		\includegraphics[width=\textwidth]{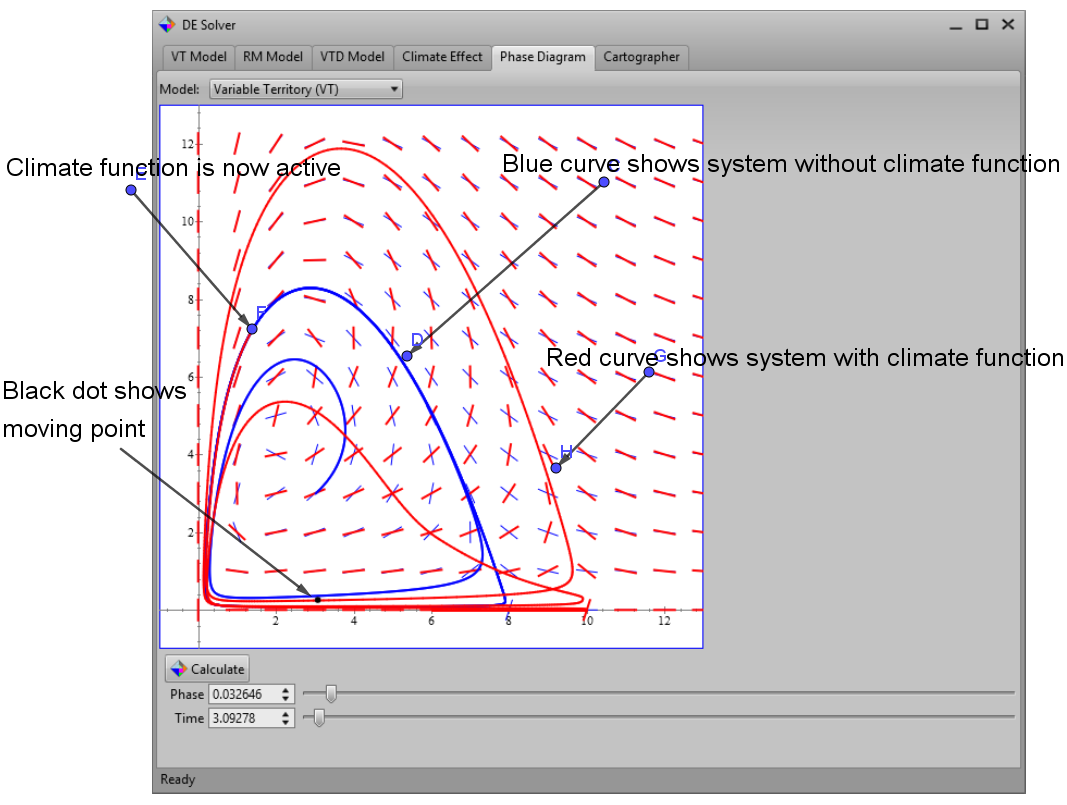}
		\caption {Screen capture of the phase diagram tool. The horizontal and vertical axes correspond respectively to the prey and predator densities}    
		\label{fig:PhaseDiagramTabScreenCapture}
	\end{subfigure}
	\hfill
	\begin{subfigure}[t]{0.37\textwidth}  
		\centering 
		\includegraphics[width=\textwidth]{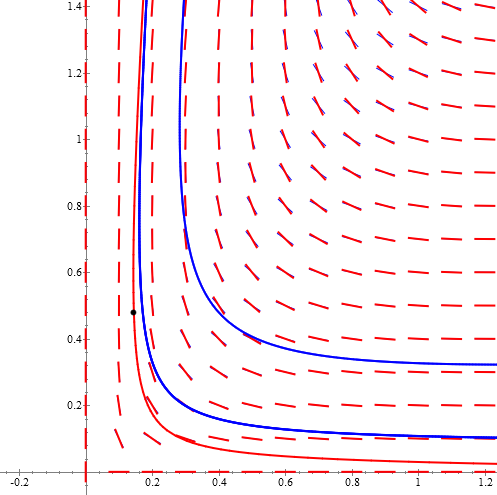}
		\caption {Close-up view showing the point where the red curve starts diverging from the blue curve, leading to extinction.}    
		\label{fig:PhaseDiagramZoomExtinction}
	\end{subfigure}
	\caption {Using phase diagrams to investigate extinction. } 
	\label{fig:PhaseDiagramTab}
\end{figure}

This new tool displays the phase diagram of the system and shows two curves (see Figure~\ref{fig:PhaseDiagramTab}): The blue curve gives the solution to~\eqref{eq:VTalleeClimate} in constant climatic conditions ($e=0$), while the red curve traces the solution from the point in time where a nonzero $e$ is introduced. The diagram also shows two vector fields: The short blue lines correspond to the $e=0$ vector field, and the short red lines correspond to the $e>0$ vector field.  Note that the blue vector field is unvarying, while the lines of the red vector field change orientation as time evolves.  

The moment when the climate function starts to affect the system is determined by the phase parameter, which can be changed interactively by moving a slider. Another slider allows the user to move a point along the red curve to see how the corresponding vector field evolves as compared to the original direction field of the autonomous system. A new solution (red curve) is calculated in real time as the user moves each slider. While the one-dimensional plot of survival time as a function of the phase (Figures~\ref{fig:SurvivalPhasePlot} and~\ref{fig:NegBumpSurvivalPhasePlot}) gives a static perspective upon survival, this tool shows its dynamic aspect. 

\begin{figure}
	\centering
	\begin{subfigure}[t]{0.45\textwidth}
		\centering
		\includegraphics[width=\textwidth]{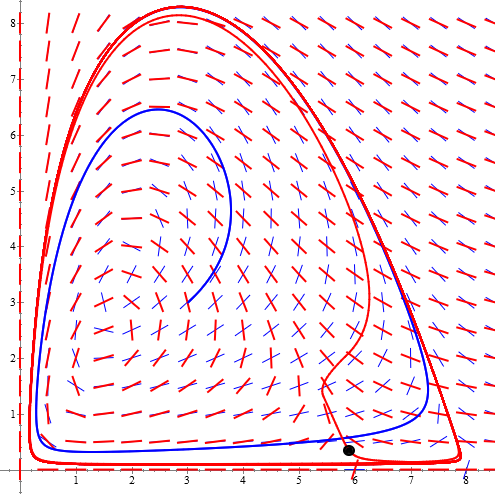}
		\caption {Case where predators survived}    
		\label{fig:PhaseDiagramSurvival}
	\end{subfigure}
	\hfill
	\begin{subfigure}[t]{0.45\textwidth}  
		\centering 
		\includegraphics[width=\textwidth]{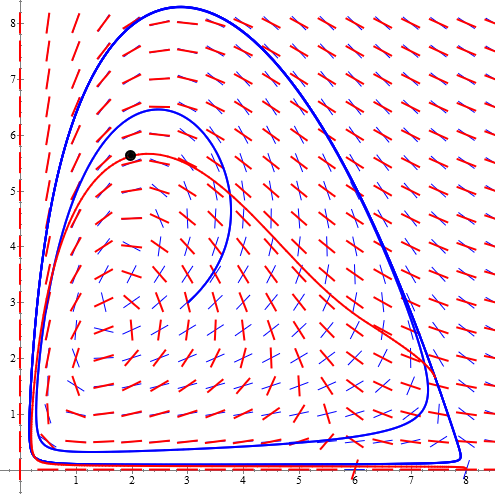}
		\caption {Case where predator went instinct.}    
		\label{fig:PhaseDiagramExtinction}
	\end{subfigure}
	\caption {Phase diagrams showing the difference between survival (a) and extinction (b) when the predator-prey system was affected by a climate function with a single negative bump of 6 years. In both cases the black dot shows the point in time corresponding to 3 years after the introduction of the climate function.} 
	\label{fig:PhaseDiagramPlots}
\end{figure}

Seen from this perspective and using a climate function with a single negative bump, we observe that survival occurs when the climate function is introduced at points in the cycle located near the bottom and bottom right of the phase diagram, that is, at points where $P$ is small, and $N$ is increasing or close to its maximum. Figure~\ref{fig:PhaseDiagramSurvival} shows a case where the climate function is introduced in this area ($phase=0.39$). The effect of the climate change is seen as the the red curve swerves to the left, then comes back in the direction of the limit circle. Note that the red curve lies on top of the blue curve for much of the limit cycle.  The same change in the climate function at a different point in the cycle ($phase=0.76$) leads to extinction as shown in Figure~\ref{fig:PhaseDiagramExtinction}. In this case, the red curve also swerves to the left, and then approaches again the limit cycle, but at a slightly lower level of predator density, causing extinction (note that in this case the red curve does not obscure the blue curve). 

Of importance is the fact that extinction occurred in this case about 11 years after the climate function was introduced. Since the negative bump had a duration of 6 years, we see that extinction happened 5 years after the negative disturbance was over. This shows that from a practical point of view it may be difficult to assess whether the species will go extinct or survive by just monitoring population densities over a period of time lasting a few years, and that the effects of a strong climate disturbance can be felt many years into the future.

\section{Discussion}
\label{sec:Discussion}

\subsection{Summary of Results}

In this work, we investigated how cyclic predator-prey dynamics might be affected by climate change, modeled as a shift in the periodicity of the switch between good years and bad years. The Predator-prey relationship was described by a VT model with Allee effects while the climate influence was modeled using climate functions that modified both the intrinsic growth rate and carrying capacity for prey. 

While this approach provided a simplified way to model climate influence, we were still confronted with many choices for describing the temporal distribution of good and bad years. This led to the creation of a software framework based on a very fast solver that would allow us to quickly test a wide range of scenarios at a high resolution of parameter values. More specifically, with the tool we created two-dimensional plots of predator survival time for each pair of parameter values. 

The central question of this investigation was to determine the conditions that could lead to extinction. We investigated both "single bump" climate functions, consisting of a single bout of $T$ "bad" years, and periodic climate functions in which the prey growth rate and carrying capacity were altered in a repeating pattern of $T_1$ "good" years (higher values of intrinsic growth rate and carrying capacity) followed by $T_2$ "bad" years (lower values of the same two parameters).

Our investigation of the single bump climate function revealed that the phase of the predator-prey limit cycle at which the climate disturbance is initiated is a critical factor in predator survival.  We found that extinction was most likely to occur if the climate disturbance was initiated at a time when prey density is very low or rapidly decreasing. 
Within the survival window (or envelope) determined by the single bump climate function, the amplitude and duration of the climate disturbance also had an important effect on predator survival.  As a rule, small amplitudes and short durations tended to promote survival. 

\subsection{Ecological Implications}

Our work indicates that cyclic predator-prey systems may be vulnerable to abrupt negative climate disturbances. In the worst case scenario, extinction of the predator can occur quickly; this outcome is expected when the onset of negative change coincides with a critical time during the cycle (when the prey density is low or is rapidly decreasing) and when the amplitude of the change and its duration are both large. Recognition of such situations could help managers assess extinction risks for vulnerable species, provided enough is known about their cyclic behaviour. 

One of the most striking and concerning aspects of the model results is the abrupt boundary between regions of parameter space where survival is assured, and regions where rapid extinction is certain.   For instance, a slight change in amplitude or duration can lead the system to extinction in a short number of years (typically less than 10). Even if the negative disturbance is overcome, that is, if the system survives the negative change, and the environment returns to normal conditions, the initial shock may still be enough to bring about extinction later on because of the earlier perturbation induced in the cycle. 

Quantifying the magnitude of the negative change that would lead to extinction is difficult at this point, but could be estimated in real biological systems given accurate parameter values. It is also possible that cyclic predator-prey systems consisting of a diverse community of predators and prey may be more resilient to the type of climate change investigated here.  More work is needed on this point.


We also recognize that the climate functions we have used are very simple.  Based on the response of the system to added noise, it is likely that our study reveals the broad patterns that can be expected in the response of predator-prey systems to a highly fluctuating climate.  More study is needed however, in designing and studying the effect of more realistic climate functions. One possibility would be to use real climate data and generate the corresponding distribution of good and bad years. For instance, we could consider that good years correspond to years where the climate indicator is close to the longtime average and bad years when the indicator departs strongly from this average. We could also experiment with distributions in which the number of good or bad years tends to increase in time, or simply follows a more random distribution. The amplitude of the climate function could be adjusted as well to follow a specific pattern. Taking advantage of the software framework we presented, many scenarios can be tested in a reasonable amount of time.

In this investigation we only considered the direct influence of climate change upon the prey intrinsic growth rate and carrying capacity. Other model parameters, however, may also be affected by climate change. One possible candidate is the predator kill rate, if changing climatic conditions make it harder (by melting snow, for example) for the predators to hunt and kill prey. By considering specific predator-prey systems for which enough data has been collected we may be able to get a better description of the complex influence that the climate can exert upon those systems.

Our study focused on the VT model with weak Allee effects and could be enlarged to take into account different models with different types of Allee effects. Our software platform was designed to accommodate different models and different climate functions. The Rosenzweig–MacArthur model has already been implemented as well as a VT model with delays. Also, more work is needed to determine good parameters for these models based on actual physical data. 

In this work we have ignored spatial dynamics.  Dispersal to more suitable habitat, however, is a well-documented response to climate change ~\citep{parmesan:2003}.  Additional work is needed connecting the type of climate change studied here with dispersal of predator-prey communities to follow range boundaries.

\section*{Acknowledgements}
This work was funded in part by NSERC DG 2016-05277 (RCT).  We also acknowledge the support of UBC Okanagan and the Institute for Biodiversity, Resilience, and Ecosystem Services (BRAES).  Finally, we are grateful to TyLab and the Computational Ecology Research Group (CERG) for many helpful discussions.

\printbibliography

\end{document}